\documentclass[10pt,letterpaper]{article}
\usepackage[top=0.85in,footskip=0.75in,marginparwidth=2in]{geometry}

\usepackage[utf8]{inputenc}

\usepackage{cite}

\usepackage{nameref,hyperref}

\usepackage{microtype}
\DisableLigatures[f]{encoding = *, family = * }

\usepackage{changepage}
\usepackage{dcolumn}   
\usepackage{bm}        
\usepackage{amssymb}   
\usepackage{amsmath}
\usepackage{mciteplus}

\usepackage[aboveskip=1pt,labelfont=bf,labelsep=period,singlelinecheck=off]{caption}

\makeatletter
\renewcommand{\@biblabel}[1]{\quad#1.}
\makeatother

\usepackage{lastpage,fancyhdr,graphicx}
\usepackage{epstopdf}
\fancyheadoffset[L]{2.25in}
\fancyfootoffset[L]{2.25in}

\usepackage{color}

\definecolor{Gray}{gray}{.25}

\usepackage{graphicx}

\usepackage{sidecap}

\usepackage{wrapfig}
\usepackage[pscoord]{eso-pic}
\usepackage[fulladjust]{marginnote}
\reversemarginpar

\begin{document}

\begin{flushleft}
{\Large
\textbf\newline{Thermodynamics of Non-linear magnetic-charged AdS black hole surrounded by quintessence, in the background of perfect fluid dark matter}
}
\newline
\\
Ragil Brand  Ndongmo\textsuperscript{1,*},
Saleh Mahamat\textsuperscript{2,5},
Thomas Bouetou Bouetou\textsuperscript{1,3},
Conrad Bertrand Tabi\textsuperscript{4},
Timoleon Crepin Kofane\textsuperscript{1,4}
\\
\bigskip
\bf{1} Department of Physics, Faculty of Science, University of Yaounde I, P.O. Box. 812, Yaounde, Cameroon,
\\

\bf{2} Department of Physics, Higher Teacher’s Training College,  University of Maroua, P.O. Box 55, Maroua, Cameroon,
\\
\bf{3} National Advanced School of Engineering, University of Yaounde I, P.O. Box. 8390, Yaounde, Cameroon,
\\
\bf{4} Department of Physics and Astronomy, Botswana International University of Science and Technology, Private Mail Bag 16, Palapye, Botswana
\\
\bf{5} The Research Center in Didactic of Fundamental and Applied Sciences, University of Maroua, Cameroon
\\
\bigskip
* nragilbrand@gmail.com

\end{flushleft}

\begin{abstract}
	In this paper, we study the thermodynamic features of a non-linear magnetic-charged AdS black hole surrounded by quintessence, in the background of perfect fluid dark matter(PFDM). After having constructed the corresponding metric, we analyse the structure of the horizon. We find that the existence of inner or outer horizon are constrained by the presence of dark matter. Afterwards, we put out the mass and the temperature of the black hole, in order to get its entropy. Subsequently, we find the expression of the pressure which leads us to get the table of critical values and the isothermal diagram. Especially, we find that the critical values of the temperature and the pressure increase as the dark matter parameter increases. Also, analysing the isothermal diagram, we observe a van der Waals-like behaviour remarked by the presence of a first-order phase transition when we cross the critical temperature. Additionally, we compute and plot the heat capacity and the Hessian matrix of the black hole mass. For the heat capacity, we find that a second-order phase transition occurs, leading the black hole to move from stable phase to unstable one. Furthermore, it comes out that this phase transition point is shifted towards higher values of the horizon radius, as we decrease the dark matter density and increase the quintessence density.
	
\end{abstract}
%

%
%




\section{Introduction}
\label{sec:intro}

The thermodynamic study of black holes is one of the most used way to apprehend black holes since the seminal works of Hawking~\cite{Hawking1974} and Bekenstein~\cite{Bekenstein1973}. Precisely, they found outstanding results such that black holes radiate as black bodies. In this way, we could find many thermodynamic quantities of the black hole, namely the temperature, the entropy, the volume, the heat capacity, and so on~\cite{bekenstein1975statistical,bekenstein1973extraction,bardeen1973four,hawking1975particle}. Since then, many works have been done on the thermodynamics of black holes~\cite{tharanath2013thermodynamics,gibbons1977cosmological,yi2011thermodynamic,ghaderi2016thermodynamics,shahjalal2019thermodynamics,Banerjee2011b,appels2016thermodynamics,Husain2009,Mahamat2018,davies1977thermodynamic,ghaffarnejad2022magnetic}.

One of the properties of black holes predicted by the Einstein theory of general relativity is known as Singularity, which is the point of space-time for which the predictive physical laws are broken down~\cite{hawking2010nature,hawking1973large}. In order to solve this problem, many alternative solutions without singularity have been constructed and they are called regular black holes. The Bardeen black hole belongs to these solutions~\cite{bardeen1968proceedings,Mahamat2018,ayon2000bardeen} with an event horizon satisfying the weak energy condition. It has been derived by introducing an energy-momentum tensor, interpreted as the gravitational field of some sort of a non-linear magnetic monopole charge $Q$. Thereby, many authors have been interested in these regular black holes and their geometrical and thermodynamic properties ~\cite{breton2005stability,abdujabbarov2013charged,ruffini2013einstein,lim2015motion,allahyari2020magnetically}.

A recent fascinating results of observational cosmology is the accelerated expansion of the Universe~\cite{riess1998observational,riess1999bvri,perlmutter1999measurements}. Moreover, this result has been confirmed by the measurement of the Cosmic Microwave Background (CMB) by the PLANCK  space Satellite~\cite{collaboration2014ade}. To explain such a phenomenon, an exotic scalar field with a large negative pressure called "dark energy" has been suggested as being the major component, up to $70\%$ of the total energy of the Universe~\cite{tharanath2013thermodynamics}. Many candidates have been proposed to be dark energy.  A well known of them is the cosmological constant. Beside it, the model used by several authors is called quintessence, which is characterised by a parameter $\epsilon$, defined as the ratio of the pressure to the energy density of the dark energy. $\epsilon$ is defined in the range $-1< \epsilon\leq -\frac{1}{3}$~\cite{shahjalal2019thermodynamics,javed2019fermions,tharanath2013thermodynamics}. Therefore, it seems interesting to study the effects of quintessence on black holes. In that way, Kiselev ~\cite{Kiselev2003} has proposed a solution corresponding to the Schwarzschild black hole surrounded by the quintessence. Afterwards, many works have been done in order to study the black hole in the quintessence field~\cite{chen2008hawking,ndongmo2021thermodynamic,yi2011thermodynamic,thomas2012thermodynamics,fernando2013nariai,ghaderi2016thermodynamics,li2014effects,xu2017kerr,younas2015strong,de2018three}.

Beside dark energy, another unsolved problem in cosmology and astrophysics is dark matter, which constitutes about 23$\%$ of the total mass-energy of the universe~\cite{xu2018kerr}, according to the Standard Model of Cosmology. Many theoretical models have been proposed to be dark matter. Beside Cold Dark Matter (CDM)~\cite{navarro1997universal}, Warm Dark Matter~\cite{dutta2021decaying,ruiz2021scalar} and Scalar Field Dark Matter~\cite{paranjape2021quantum,padilla2021consequences}, the Perfect fluid dark matter(PFDM) is one among them, and it has been shown that the PFDM can explain the asymptotically flat rotation curves concerning spiral galaxies~\cite{siddhartha2003quintessence}. Hence the introduction of PFDM in many works concerning black holes~\cite{saurabh2021imprints,shaymatov2021testing,ghosh2021charged,ma2021shadow}. 

AdS spacetime is remarked by a constant negative scalar curvature, and corresponds to a negative cosmological constant $\Lambda$ on the spacetime, with a positive pressure. Since the introduction of AdS spacetime, it has been proved that cosmological constant can have an influence in high energy astrophysical objects, such as active galactic nuclei and supermassive black holes\cite{stuchlik2005influence}. Hence, it can be interesting to study the impact of dark energy or dark matter onto the behaviour of AdS black holes, as it has been done in\cite{xu2017kerr,sadeghi2020ads,javed2019fermions,cao2021joule,xu2018kerr,xu2019perfect,ahmed2019effect}. Furthermore, it has also been studied the quintessence AdS black hole in the framework of holography\cite{chen2013holographic}.

Since seminal work of Hawking and Page~\cite{Hawking1983}, it has been shown that black holes undergo to a phase transition, in the AdS/CFT correspondence. Furthermore, the understanding of the phase transition could be extended to the one between small-large black hole, as in~\cite{farhangkhah2021extended,wu2021ruppeiner,hendi2021critical}, for which they showed a complete analogy with the van der Waals liquid-gas system. Unlike the classical thermodynamics, there is no usual $P-V$ term in the first law of the black hole thermodynamics. Therefore, in order to restore it, it has been suggested that the Cosmological constant $\Lambda$ plays the role of the pressure $P$ and its conjugate quantity as a thermodynamic volume $V$ in the extended phase space, and then the black hole mass considered as the enthalpy, as suggested in~\cite{kastor2009enthalpy,kastor2011mass,kastor2018black}. Thereby, this reasoning has enriched several studies on the thermodynamic study of black holes~\cite{dolan2011pressure,dolan2011cosmological,hendi2020instability,HONG2019114826,he2019weak,toledo2020kerr,guo2020continuous,chabab2020thermodynamic,liang2021thermodynamics}.

Another way to study the black hole phase transitions is through the behaviour of its heat capacity~\cite{tharanath2013thermodynamics}. Especially, Husain and Mann~\cite{Husain2009} suggested that the specific heat of a black hole becomes positive after a phase transition near the Planck scale, and the presence of a discontinuity in the plot of the heat capacity shows the presence of a second-order phase transition. Afterwards, it has been studied in several works, in order to explore the black hole phase transition(see~\cite{rodrigue2020thermodynamic,cai2009thermodynamics,Mahamat2018,rodrigue2018thermodynamics,tharanath2013thermodynamics,tharanath2014phase,li2020thermodynamic}). Especially, Nam~\cite{nam2018on} derived a non-linear magnetic-charged black hole surrounded by quintessence, and studied its thermodynamic stability. As a result, he found that the black hole may undergo, at a critical temperature, a thermal phase transition, between a larger unstable black hole and a smaller stable black hole. Therefore, what could we have if we also take into account the presence of dark matter, in the AdS space-time?

In this paper,
we aim at studying the impact of these two quantities on the thermodynamic behaviour of the non-linear magnetic-charged AdS black hole.

In an effort to find possible solutions to this concern, the paper is organised as follows. In Section~\eqref{sec:metric}, we derive the metric corresponding to the non-linear magnetic-charged AdS black hole surrounded by quintessence, in the background of perfect fluid dark matter. Afterwards, in section~\eqref{sec:horizon1}, we study the horizon structure through the event horizon property. In Section~\eqref{sec:thermo}, by considering the cosmological constant acting as a dynamical pressure, we study the thermodynamic stability and the phase transitions of the black hole, and we put out the effects of the quintessence energy and the PFDM in the background of the non-linear magnetic-charged AdS black hole. The Section~\eqref{sec:concl}  is devoted to the conclusion.

\section{\label{sec:metric} Non-linear magnetic-charged AdS black hole surrounded by quintessence and the perfect fluid dark matter}

In the presence of quintessence, the action corresponding to the Einstein gravity coupled to a non-linear electromagnetic field in the four-dimensional AdS space-time  and in the presence of PFDM can be expressed as~\cite{li2012galactic,xu2018kerr,xu2019perfect,sadeghi2020universal,salazar1987duality,novello2000singularities,nam2020higher}

\begin{eqnarray}\label{act}
\begin{array}{r c l}
S&=&\int d^4x\sqrt{-g}\left[\frac{c^4}{16\pi G}(R-2\Lambda)-(\mathcal{L}_\textmd{charge}\right.\\ 
&+&\left.4\pi\mathcal{L}_\textmd{PFDM}-\mathcal{L}_\textmd{quint})\right]
\end{array}
\end{eqnarray}

where $R$ is the scalar curvature, $\Lambda$ is the cosmological constant, $\mathcal{L}_\textmd{charge}$ is the non-linear electrodynamic term and is a function of the invariant $F_{\mu\nu}F^{\mu\nu}/4\equiv F$, with $F_{\mu\nu}=\partial_\mu A_\nu-\partial_\nu A_\mu$, being the Faraday tensor of electromagnetic field, and $A_\nu$ is the gauge potential of the electromagnetic field, $g$ is the determinant of the metric tensor $g_{\mu\nu}$, $G$ is the Newton gravity constant and $c$ is the light speed. The expression $\mathcal{L}_\textmd{charge}$ is given by~\cite{nam2018on,nam2018non,chen2020optical,ma2021shadow}
\begin{equation}\label{L_F}
\mathcal{L}_\textmd{charge}=\frac{3M}{|Q|^3}\frac{(2Q^2F)^{3/2}}{\left[1+(2Q^2F)^{3/4}\right]^2}    \ ,
\end{equation}

where $M$ and $Q$ are the parameters associated with mass and magnetic charge of the system, respectively.

The PFDM term in the action (\ref{act}) is remarked by the term $\mathcal{L}_\textmd{PDFM}$, which is the PFDM Lagrangian density, and $\mathcal{L}_\textmd{quint}$ is the term corresponding to quintessence, which is given by~\cite{sadeghi2020ads,ghosh2018lovelock,bohmer2015interacting}
\begin{equation}
\mathcal{L}_\textmd{quint}=-\frac{1}{2}(\nabla\phi)^2-V(\phi),
\end{equation}
where $\phi$ is the quintessential scalar field, and $V(\phi)$ is the potential.



Therefore, applying variational principle from Eq. (\ref{act}), meaning that making the  extremization of the action, with respect to the inverse of the metric $g^{\mu\nu}$, 
the action leads to

\begin{eqnarray}
\label{S1}    0&=&\frac{1}{\sqrt{-g}}\frac{\delta S(\textmd{action})}{\delta g^{\mu\nu}}\\ 
\label{S2}     &=&\frac{1}{\sqrt{-g}}\left\{ \frac{c^4}{16\pi G}\int d^4x \frac{\delta (\sqrt{-g}(R-2\Lambda))}{\delta g^{\mu\nu}} \right.\\
\label{S3}     &-&\left(\int d^4x \frac{\delta (\sqrt{-g}\mathcal{L}_\textmd{charge})}{\delta g^{\mu\nu}}\right.\\
\label{43}	   &+&\int d^4x 4\pi\frac{\delta (\sqrt{-g}\mathcal{L}_\textmd{PFDM})}{\delta g^{\mu\nu}}\\
\label{S5}     &-& \left.\left.\int d^4x \frac{\delta (\sqrt{-g}\mathcal{L}_\textmd{quint})}{\delta g^{\mu\nu}}\right)\right\}.
\end{eqnarray}


Having this in mind, one can get
\begin{equation}\label{Guv-2}
\begin{array}{r c l}
G_{\mu\nu}+\Lambda g_{\mu\nu}&=&\frac{8\pi G}{c^4}\left(T_{\mu\nu}(\textmd{charge})+4\pi T_{\mu\nu}(\textmd{PFDM})\right.\\
&-&\left.T_{\mu\nu}(\textmd{quint})\right)
\end{array}
\end{equation}

In Eq. (\ref{Guv-2}), the different energy-momentum tensors are expressed as follows\cite{nam2018non,gonzalez2008exact,bohmer2015interacting}

\begin{eqnarray}
\label{Tuv-3}   T_{\mu\nu}(\textmd{charge})=& \frac{2\delta (\sqrt{-g}\mathcal{L}_\textmd{charge})}{\delta g^{\mu\nu}}\\
=&-\frac{\partial \mathcal{L}_\textmd{charge}}{\partial F}F_\mu^\beta F_{\nu\beta}+\mathcal{L}_\textmd{charge}g_{\mu\nu},\\ 
\label{Tuv-2}   T_{\mu\nu}(\textmd{quint}) =& \frac{2\delta (\sqrt{-g}\mathcal{L}_\textmd{quint})}{\delta g^{\mu\nu}}\\
=& \left[\nabla_\mu\phi\nabla_\nu\phi-\frac{1}{2}g_{\mu\nu}\left((\nabla\phi)^2+2V(\phi)\right)\right],\\
\label{Tuv-1}   T_{\mu\nu}(\textmd{PFDM})  =& \frac{2\delta (\sqrt{-g}\mathcal{L}_\textmd{PFDM})}{\delta g^{\mu\nu}}.
\end{eqnarray}

In units with the normalization of Newton gravity constant $G$, light speed $c$ and the number $\pi$ by $\frac{4\pi G}{c^4}=1$, we finally get the Einstein-Maxwell equations of motion, expressed in the contravariant coordinates as follows

\begin{eqnarray}
\begin{array}{r c l}
\label{Guv}   G^\nu_\mu+\Lambda \delta^\nu_\mu&=&2\left(\frac{\partial\mathcal{L}_\textmd{charge}}{\partial F}F_{\mu\rho}F^{\nu\rho}-\delta^\nu_\mu\mathcal{L}_\textmd{charge}\right.\\
&+&\left.4\pi T^\nu_\mu(\textmd{PFDM})-T^\nu_\mu(\textmd{quint})\right),
\end{array}
\end{eqnarray}

\begin{eqnarray}
\label{Max1}  \nabla_\mu\left(\frac{\partial\mathcal{L}_\textmd{charge}}{\partial F}F^{\nu\mu}\right)&=0,\\
\label{Max2}  \nabla_\mu*F^{\nu\mu}&=0.
\end{eqnarray}

Since dark matter is considered as a kind of perfect fluid, the energy-momentum tensor is then written as $T_\mu^\nu=\textmd{diag}[-\rho,p,p,p]$~\cite{li2012galactic,xu2018kerr,xu2019perfect}, with $\rho$ and $p$ being the energy density and the pressure, respectively. Furthermore, in the simplest case, we assume 	$\frac{p}{\rho}=\delta-1$, where $\delta$ is a constant~\cite{li2012galactic}.

Now, since we need to find a spherically symmetric AdS black hole solution of the mass $M$ and the magnetic charge $Q$ in the quintessence and PFDM, the metric has to be written with ansatz~\cite{nam2018on,xu2018kerr,rizwan2020coexistent}
\begin{eqnarray}\label{metric}
\begin{array}{r c l}
ds^2&=&-e^\nu dt^2+e^\lambda dr^2+r^2(d\theta^2+\sin^2\theta d\phi^2)\\ 
&=&-f(r)dt^2+\frac{1}{f(r)}dr^2+r^2(d\theta^2+\sin^2\theta d\phi^2),\\
&\textnormal{with}&\ \ f(r)=1-\frac{2m(r)}{r}-\frac{\Lambda}{3}r^2.
\end{array}
\end{eqnarray}
The ansatz we use for the Maxwell field is expressed as~\cite{nam2018on,xu2018kerr,ayon2000bardeen}
\begin{equation}
F_{\mu\nu}=\left(\delta_\mu^\theta\delta_\nu^\varphi-\delta_\nu^\theta\delta_\mu^\varphi\right)B(r,\theta).
\end{equation}

Here, one can notice that the magnetic charge $Q$ is defined as~\cite{nam2018on} 

\begin{equation}\label{sphere}
\frac{1}{4\pi}\int_{S_2^\infty}\textbf{\textit{F}}=Q,
\end{equation}
with $S_2^\infty$ being a two-sphere at the infinity. Note furthermore that  $Q$ is the integral constant which should be used to integrate Eqs. (\ref{Max1}) and (\ref{Max2}) in order to find the expression of $F_{\mu\nu}$, and $M$ is the one which will allow us to find the expression of $m(r)$. Now, taking into account Eqs. (\ref{Max1}), (\ref{Max2}) and (\ref{sphere}), we have~\cite{nam2018on}

\begin{equation}\label{B}
B(r,\theta)=Q\sin(\theta),
\end{equation}
which leads straightforwardly to
\begin{equation}\label{F}
F=\frac{Q^2}{2r^4}.
\end{equation}
Now, replacing it into Eq. (\ref{L_F}), we get the non-linear electrodynamic term as
\begin{equation}
\mathcal{L}_\textmd{charge}=\frac{3MQ^3}{(r^3+Q^3)^2}.
\end{equation}

Considering the time component of Eq. (\ref{Guv}), we get
\begin{eqnarray}
\begin{array}{r c l}
\label{Guv1}     G^t_t+\Lambda \delta^t_t&=&2\left(\frac{\partial\mathcal{L}_\textmd{charge}}{\partial F}F_{t\rho}F^{t\rho}-\delta^t_t\mathcal{L}_\textmd{charge}\right.\\
&+&\left.4\pi T^t_t(\textmd{PFDM})-T^t_t(\textmd{quint})\right).
\end{array}
\end{eqnarray}
Now, since the time components of energy-momentum for the PFDM and for quintessence are related to their energy density, they are expressed as follows~\cite{zhang2021regular,Kiselev2003}

\begin{eqnarray}\label{Tuv1}
T^t_t(\textmd{PFDM})&=&\frac{1}{8\pi}\frac{\alpha}{r^3},\\ 
T^t_t(\textmd{quint})&=&-\frac{3\epsilon c_q}{2r^{3(\epsilon+1)}},
\end{eqnarray}
where $\alpha$ denotes the intensity of the PFDM, $c_q$ and $\epsilon$ are the quintessence parameters.

Therefore, from Eq. (\ref{Guv1}), we get straightforwardly
\begin{equation}\label{Guv2}
G_t^t+\Lambda=-2\mathcal{L}_\textmd{charge}+\frac{\alpha}{r^3}+\frac{3\epsilon c_q}{r^{3(\epsilon+1)}}, 
\end{equation}
since $\delta_t^t=1$, and $F_{t\rho}F^{t\rho}=0$.

Before solving Eq. (\ref{Guv2}), we first have to find the component $G_{tt}$ of the Einstein tensor, which is obtained using the metric~(\ref{metric}), by

\begin{equation}\label{Guv3}
G_{tt}=e^\nu\left[\frac{1}{r^2}-e^{-\lambda}\left(\frac{1}{r^2}-\frac{\lambda'}{r}\right)\right].
\end{equation}
Developing it, we get
\begin{eqnarray*}
G_{tt}&=&f(r)\left[\frac{1}{r^2}-f(r)\left(\frac{1}{r^2}-\frac{1}{r}\left(-\frac{f'(r)}{f(r)}\right)\right)\right] \\
&=& f(r)\left[\frac{1}{r^2}-f(r)\left(\frac{1}{r^2}-\frac{1}{rf(r)}\left[2\left(\frac{1}{r}\frac{dm(r)}{dr}\right.\right.\right.\right.\\
&-&\left.\left.\left.\left.\frac{m(r)}{r^2}+\frac{\Lambda}{3}r^2\right)\right]\right)\right]\\
&=&f(r)\left(\frac{2}{r^2}\frac{dm(r)}{dr}+\Lambda\right).
\end{eqnarray*}

with $\lambda=-\ln f(r)$

Therefore, the component $G_t^t$ of Eq. (\ref{Guv3}) is expressed as
\begin{equation}
G_t^t=g^{tt}G_{tt}=(-f(r))^{-1}f(r)\left(\frac{2}{r^2}\frac{dm(r)}{dr}+\Lambda\right),
\end{equation}
meaning that
\begin{equation}\label{Guv4}
G_t^t=-\frac{2}{r^2}\frac{dm(r)}{dr}-\Lambda.
\end{equation}

Now, replacing Eq. (\ref{Guv4}) into Eq. (\ref{Guv2}), we get 
\[
\frac{dm(r)}{dr}=\frac{3MQ^3r^2}{(r^3+Q^3)^2}-\frac{\alpha}{2r}-\frac{3\epsilon c_q}{2r^{3\epsilon+1}}.
\]
By integration, we obtain
\begin{equation}
m(r)=-\frac{MQ^3}{r^3+Q^3}-\frac{\alpha}{2}\ln\frac{r}{|\alpha|}+\frac{c_q}{2r^{3\epsilon}}+C^{st}.
\end{equation}
Now, to find the integral constant $C^{st}$, we will use the boundary condition~\cite{nam2018on,zhang2021regular} 
\begin{eqnarray*}
M&=&\lim\limits_{r\rightarrow \infty}\left\{m(r)+\frac{\alpha}{2}\ln\frac{r}{|\alpha|}-\frac{c_q}{2r^{3\epsilon}}\right\}\\
&=&\lim\limits_{r\rightarrow \infty}\left\{-\frac{MQ^3}{r^3+Q^3}+C^{st}\right\},
\end{eqnarray*}

leading to
\begin{equation}
C^{st}=M.
\end{equation}

Hence, the mass $m(r)$ and the function $f(r)$ are respectively given by
\begin{equation}
m(r)=\frac{Mr^3}{r^3+Q^3}-\frac{\alpha}{2}\ln\frac{r}{|\alpha|}+\frac{c_q}{2r^{3\epsilon}},
\end{equation}

\begin{equation}
f(r)=1-\frac{2Mr^2}{r^3+Q^3}+\frac{\alpha}{r}\ln\frac{r}{|\alpha|}-\frac{c_q}{r^{3\epsilon+1}}-\frac{\Lambda}{3}r^2.
\end{equation}

Hereby, the spherically symmetric solution for the action (1) is obtained as

\begin{equation}
\label{metric1}
ds^2=-f(r)dt^2+\frac{1}{f(r)}dr^2+r^2(d\theta^2+\sin^2\theta d\phi^2),
\end{equation}

\begin{equation}
\textnormal{with}\ f(r)=1-\frac{2Mr^2}{r^3+Q^3}+\frac{\alpha}{r}\ln\frac{r}{|\alpha|}-\frac{c_q}{r^{3\epsilon+1}}-\frac{\Lambda}{3}r^2.
\end{equation}
Let us notice first that if we replace the quintessence parameter $c=0$ and the cosmological constant $\Lambda=0$ into Eq. (\ref{metric1}), we recover the metric of non-linear magnetic-charged  black hole surrounded by dark matter considered by Ma et \textit{al.}~\cite{ma2021shadow}.

\section{Horizon structure analysis}\label{sec:horizon1}
In order to study the event horizon of the black hole, we need to use the following horizon property~\cite{benavides2020rotating,nam2018on,tharanath2014phase} 
\begin{equation}
g^{rr}=0.
\end{equation}
Therefore, through this equation, we obtain

\begin{equation}\label{f(r)1}
f(r)=1-\frac{2Mr^2}{r^3+Q^3}+\frac{\alpha}{r}\ln\frac{r}{|\alpha|}-\frac{c_q}{r^{3\epsilon+1}}-\frac{\Lambda}{3}r^2=0.
\end{equation}

%
This means that the horizons of the  black hole are solutions of  Eq. (\ref{f(r)1}). In the literature, it has been shown that a charged black hole with quintessence has possibly three horizons: the inner horizon $r_-$, the event horizon $r_+(\geq r_-)$ and the quintessence horizon $r_q(\geq r_+)$\cite{nam2018on}.

\begin{figure*}
\centering
\begin{minipage}[!h]{7cm}
	\centering
	\includegraphics[scale=0.30]{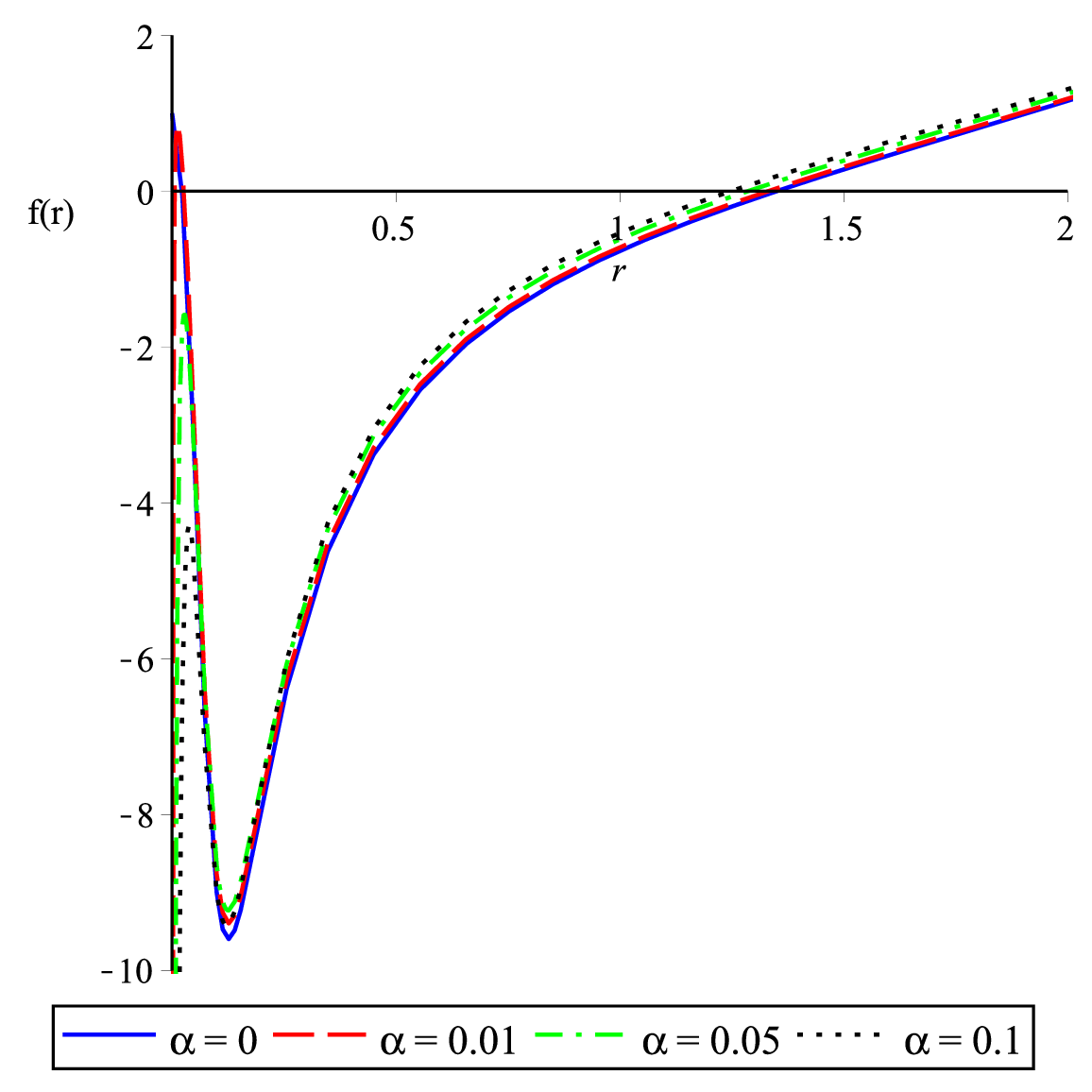}
	
	(a) 
\end{minipage}
\begin{minipage}[!h]{7cm}
	\centering
	\includegraphics[scale=0.30]{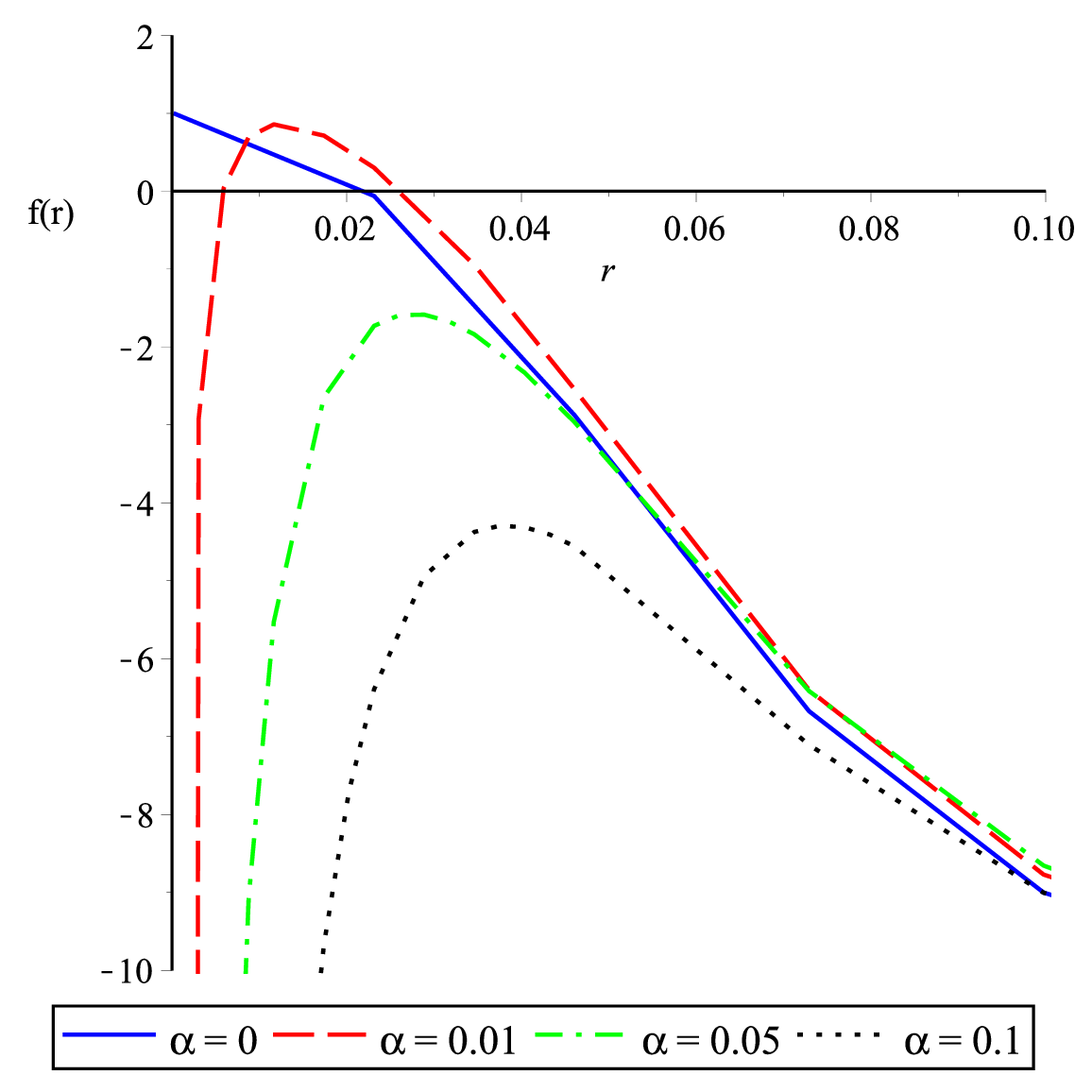}
	
	(b)
\end{minipage}
\caption{\label{hori}Variation of the function $f(r)$ in term of radius with  $(Q, c_q, \epsilon,\Lambda)=(0.1, 0.1, -0.6,-1)$.} 
\end{figure*}

The variation of $f(r)$ is depicted in Fig. \eqref{hori}, in term of the radius, for different values of the dark matter parameter $\alpha$. This plot, especially Fig. \eqref{hori} (a), shows that the black hole has a quintessence horizon, which is in our case, in the range $[1,1.5]$. However the black  hole cannot necessarily have the inner and outer horizons, as it is shown in Fig. \eqref{hori}(b). Indeed, this figure shows that in the absence of dark matter, it only has an outer horizon(for $\alpha=0$). Moreover, in the case of very small dark matter parameter(for example $\alpha=0.01$), we see the presence of two horizons, but for higher values of dark matter parameter(for example $\alpha=0.05$ or $\alpha=0.1$) there is no horizon, which leads the black hole to have only the quintessence horizon. This analysis tells us that dark matter has a considerable effect on the behaviour of the non-linear magnetic-charged black hole horizon.


Now, using the horizon property~\cite{nam2018on,tharanath2014phase}, and solving the following equation at the horizon

\begin{equation}
f(r_h)=0,
\end{equation}
leads to
\begin{equation}\label{M1}
M=\frac{(r_h^3+Q^3)}{2r_h^2}\left(1+\frac{\alpha}{r_h}\ln\frac{r_h}{|\alpha|}-\frac{c_q}{r_h^{3\epsilon+1}}-\frac{\Lambda}{3}r_h^2\right).
\end{equation}

Eq. (\ref{M1}) gives the relation between the black hole mass and its horizon radius.

\section{Thermodynamic phase transition}
\label{sec:thermo}

For the next step, we will make the thermodynamic analysis of the black hole. First, we will get the isothermal $P-r_h$ diagram, and to do that, we consider that the cosmological constant term $\Lambda$ will act as a dynamical pressure. Thus, we could write~\cite{dolan2011pressure,dolan2011cosmological,hendi2020instability,HONG2019114826,he2019weak,toledo2020kerr,guo2020continuous,chabab2020thermodynamic,liang2021thermodynamics}
\begin{equation}\label{P1}
P=-\frac{\Lambda}{8\pi}.
\end{equation} 
Therefore, substituting $\Lambda$ from Eq. (\ref{P1}) into Eq. (\ref{M1}), we readily find 

\begin{equation}\label{M2}
M=\frac{(r_h^3+Q^3)}{2r_h^2}\left(1+\frac{\alpha}{r_h}\ln\frac{r_h}{|\alpha|}-\frac{c_q}{r_h^{3\epsilon+1}}+\frac{8\pi P}{3}r_h^2\right).
\end{equation}

The Hawking temperature is found through the surface gravity definition at the horizon~\cite{nam2018on}
\begin{eqnarray}\label{T1}
\begin{array}{r c l}
T&=& \frac{\kappa}{2\pi}=\frac{f'(r_h)}{4\pi}\\
&=&\frac{1}{4\pi}\left\{-2M\left[\frac{2r_h(r_h^3+Q^3)-3r^4}{(r_h^3+Q^3)^2}\right]-\frac{\alpha}{{r_h}^{2}}\,\ln  \left. {\frac {r_h}{ \left| \alpha\right| }} \right. \right.\\
&+&{\frac {\alpha}{{r_h}^{2}}}+\left.{\frac {c_q\left( 3\,\epsilon +1 \right) }{{r_h}^{3\,\epsilon +2}}}+\frac{16\pi P}{3}r_h\right\}.
\end{array}
\end{eqnarray}

Thus, using Eq. (\ref{M2}), Eq. (\ref{T1}) may become as
\begin{eqnarray}\label{T2}
\begin{array}{r c l}
T&=& \frac{1}{4\pi (r_h^3+Q^3)}\left[\frac{r_h^3-2Q^3}{r_h}+\frac{3c_q\epsilon}{r_h^{3\epsilon+2}}\left(r_h^3+Q^3\left(\frac{\epsilon+1}{\epsilon}\right)\right)\right.\\
&+& \left.\frac{\alpha Q^3}{r_h^2}\left(1-3\ln\frac{r_h}{|\alpha|}\right)+\alpha r_h+8\pi Pr_h^4\right].
\end{array}
\end{eqnarray}

Eq. (\ref{T2}) represents the Hawking temperature of the non-linear magnetic-charged black hole in the quintessence field, surrounded by perfect fluid dark matter.

Let us notice that for $\alpha \approx 0$ and $P=0$, meaning that in the absence of both dark matter and cosmological constant, we find the Hawking temperature of the non-linear magnetic-charged black hole surrounded by quintessence which has been obtained by Nam~\cite{nam2018on}, given by
\begin{equation}
T= \frac{1}{4\pi (r_h^3+Q^3)}\left[\frac{r_h^3-2Q^3}{r_h}+\frac{3c_q\epsilon}{r_h^{3\epsilon+2}}\left(r_h^3+Q^3\left(\frac{\epsilon+1}{\epsilon}\right)\right)\right].
\end{equation} 
The entropy is obtained from the temperature and mass, through the relation
\begin{equation}\label{entro1}
S=\int\frac{1}{T}\frac{\partial M}{\partial r_h}dr_h.
\end{equation}


From Eq. (\ref{M2}), we have obtained the first derivative of the mass with respect to the horizon radius as

\begin{eqnarray}\label{M/r}
\begin{array}{r c l}
\frac{\partial M}{\partial r_h}&=& \frac{1}{2r^2}\left[\frac{r_h^3-2Q^3}{r_h}+\frac{3c_q\epsilon}{r_h^{3\epsilon+2}}\left(r_h^3+Q^3\left(\frac{\epsilon+1}{\epsilon}\right)\right)\right.\\
&+&\left.\frac{\alpha Q^3}{r_h^2}\left(1-3\ln\frac{r_h}{|\alpha|}\right)+\alpha r_h+8\pi Pr_h^4\right].
\end{array}
\end{eqnarray}

Hence, from Eq. (\ref{T2}) and (\ref{M/r}), the entropy of the black hole in (\ref{entro1}), is straightforwardly obtained  as
\begin{equation}\label{entro2}
S=\pi r_h^2\left(1-\frac{2Q^3}{r_h^3}\right).
\end{equation}

Now, the next step consists of deriving the expression of the pressure $P$ as a function of $T$ and $r_h$, then ending by analysing the critical behaviour of $P-r_h$ diagram.


From Eq. (\ref{T2}), one can explicitly express the pressure $P$ as a function of $T$ and $r_h$ as
\begin{eqnarray}
\begin{array}{r c l}
P&=&\frac{1}{8\pi r_h^4}\left[ 4\pi(r_h^3+Q^3)T-\frac{r_h^3-2Q^3}{r_h}\right.\\
&-&\left.\frac{3c_q\epsilon}{r_h^{3\epsilon+2}}\left(r_h^3+Q^3\left(\frac{\epsilon+1}{\epsilon}\right)\right)-\frac{\alpha Q^3}{r_h^2}\left(1-3\ln\frac{r_h}{|\alpha|}\right)\right.\\
&-&\left.\alpha r_h \right].
\end{array}
\end{eqnarray}

Since the black hole mass $M$ is most naturally associated with the enthalpy $H$ of the black hole in the extended phase space~\cite{tharanath2014phase}, the expression of the volume can be expressed as follows
\begin{equation}
V=\left(\frac{\partial H}{\partial P}\right)_{r_h,Q}=\left(\frac{\partial M}{\partial P}\right)_{r_h,Q}=\frac{4}{3}\pi(r^3_h+Q^3).
\end{equation}

However, given that the expression of pressure would be less complex if we express it in term of horizon radius rather than volume, we will make our thermodynamic analysis with the isothermal $P-r_h$ diagram. Therefore, we putted out in table~\ref{tab1} critical values which leads us to have an inflexion point in the isothermal $P-r_h$ diagram. They are found through the following system of equations
\begin{equation}\label{iso1}
\left(\frac{\partial P}{\partial r_h}\right)_T=0, \ \ \left(\frac{\partial^2 P}{\partial r_h^2}\right)_T=0.
\end{equation}
Next, we investigate Eq. (\ref{iso1}) numerically in order to get the Table~\ref{tab1}, since it is not a trivial task to solve it analytically.

Therefore, it is shown in the present calculation(see table~\ref{tab1}) the impact of PFDM on the critical values. remarkably, we can notice that when increasing the PFDM parameter $\alpha$, we see that $r_{h_c}$ increases for $\alpha<0.2$ and then decreases when $\alpha>0.2$. At the same time, we notice that $T_c$ and $P_c$ increase as we increase $\alpha$.

\begin{table}
\begin{center}
	\begin{tabular}{|c c c c |}
		\hline
		$\alpha$	& $r_{h_c}$   &  $T_{c}$   & $P_c$    \\
		\hline
		0.01 & \ \ 2.7109 \ \  & \ \ 0.0126\  \  & \  \  0.0032   \\
		
		0.05 & \ \ 2.7546 \ \  & \ \ 0.0127\  \  & \  \  0.0032   \\
		
		0.1  & \ \ 2.7756 \ \  & \ \ 0.0132\  \  & \  \  0.0033   \\
		
		0.2  & \ \ 2.7763 \ \  & \ \ 0.0148\  \  & \  \  0.0034   \\
		
		0.4  & \ \ 2.7090 \ \  & \ \ 0.0190\  \  & \  \  0.0039   \\
		
		0.6  & \ \ 2.5947 \ \  & \ \ 0.0240\  \  & \  \  0.0045   \\	  
		
		0.8  & \ \ 2.4418 \ \  & \ \ 0.0308\  \  & \  \  0.0054   \\			  
		\hline
	\end{tabular}
\end{center}
\caption[Un tableau] {\label{tab1} Critical values for $(Q, c_q, \epsilon)=(1, 0.2, -2/3)$ for different dark matter parameters $\alpha$.}
\end{table}

\begin{figure*}
\centering
\begin{minipage}[!h]{7cm}
	\centering
	\includegraphics[scale=0.40]{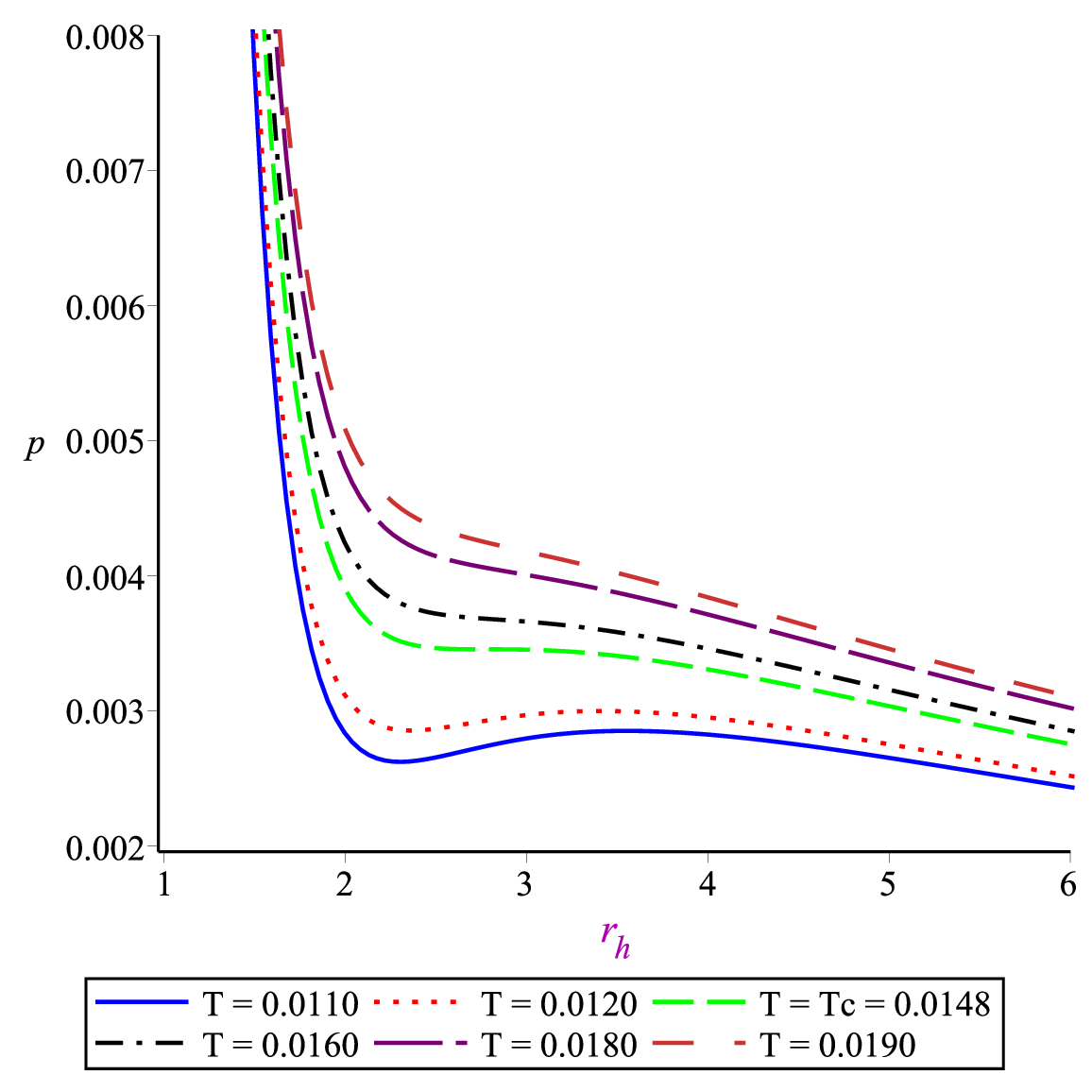}
	
	(a) For $\alpha=0.2.$
\end{minipage}
\begin{minipage}[!h]{7cm}
	\centering
	\includegraphics[scale=0.40]{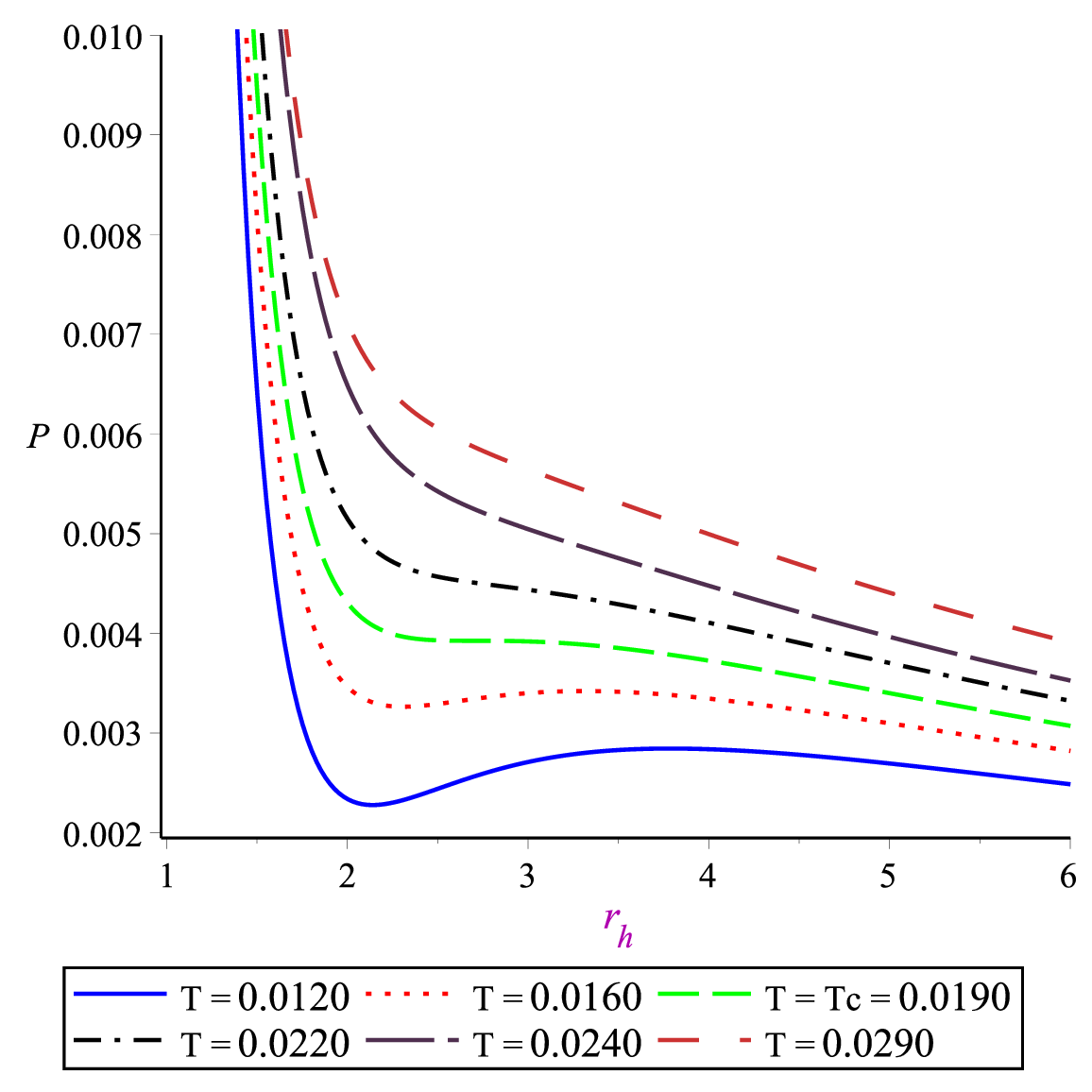}
	
	(b) For $\alpha=0.4.$
\end{minipage}
\begin{minipage}[!h]{7cm}
	\centering
	\includegraphics[scale=0.40]{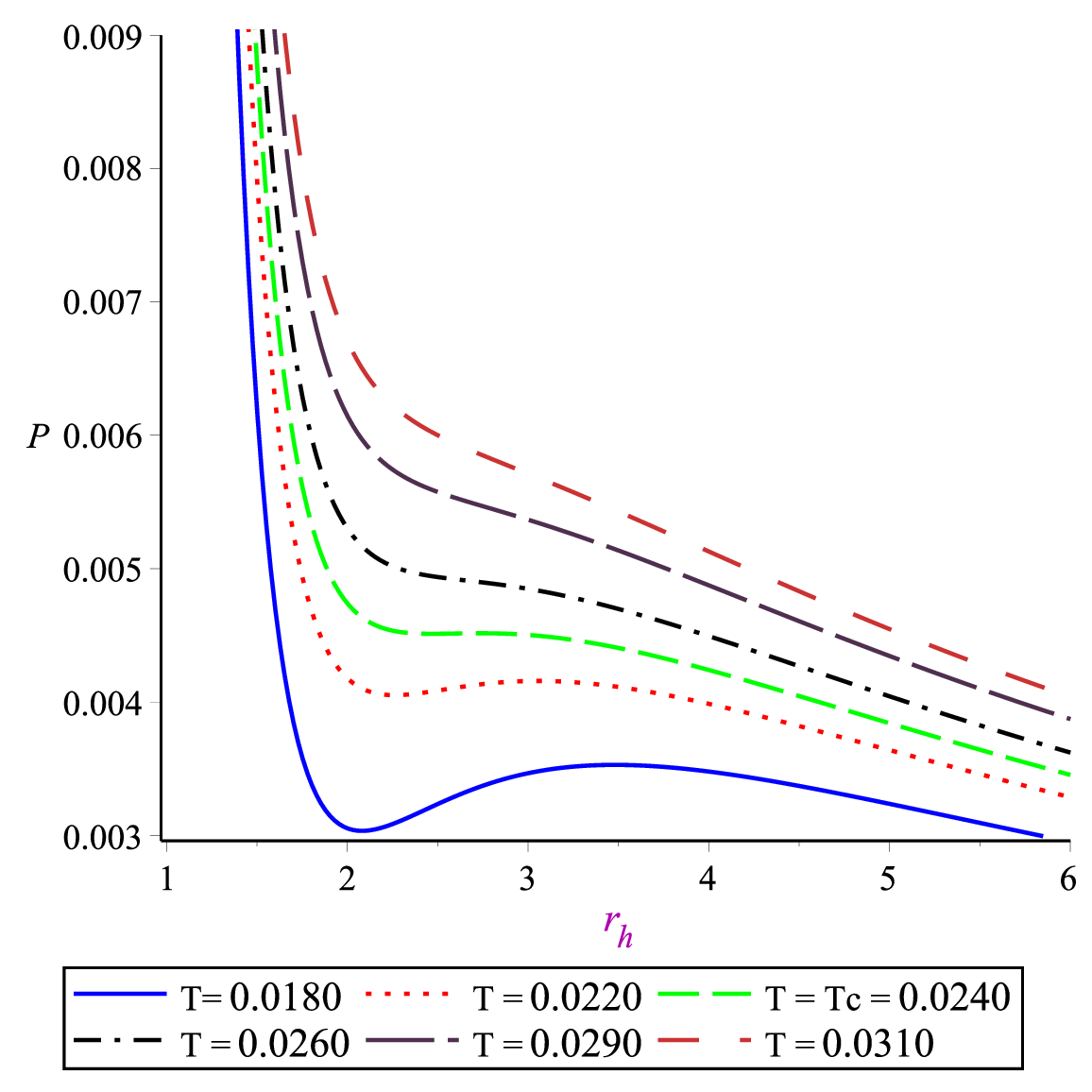}
	
	(c) For $\alpha=0.6.$
\end{minipage}
\begin{minipage}[!h]{7cm}
	\centering
	\includegraphics[scale=0.40]{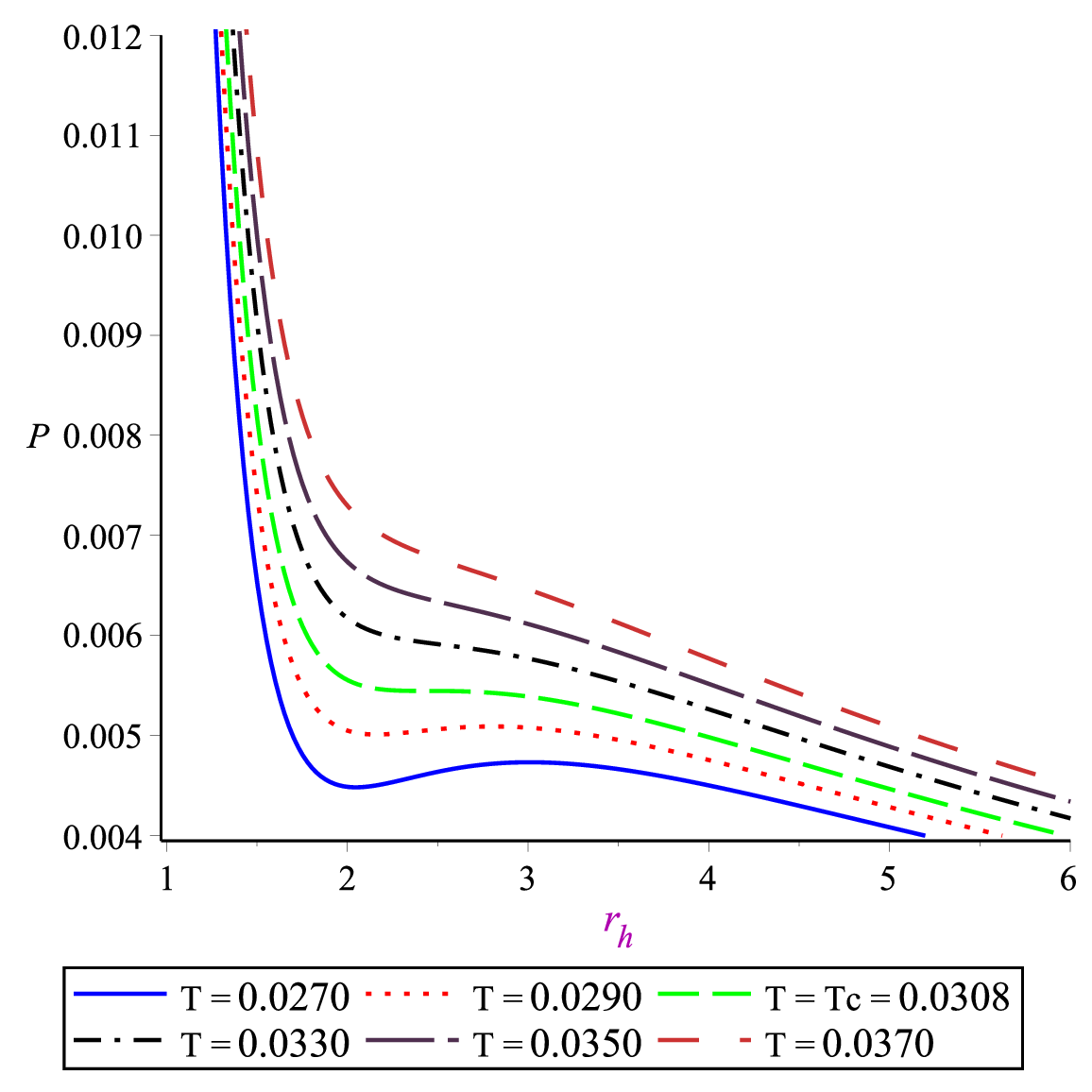}
	
	(d) For $\alpha=0.8.$
\end{minipage}
\caption{\label{fig1}Variation of pressure for different values of $\alpha$, with $(Q, c_q, \epsilon)=(1, 0.2, -2/3)$.} 
\end{figure*}


In Fig.~\ref{fig1}, we plotted the $P-r_h$ diagram. Analysing this plot, we observe a van der Waals like behaviour. This means that a first-order phase transition occurs, moving from $T < T_c$ to $T > T_c$. Here, the  first region($T < T_c)$ is remarked by one value of the horizon radius for a high pressure, and two or three horizon radii for low pressure, and the second region $(T > T_c)$ is remarked by only one horizon radius, for any value of pressure $P$. Furthermore, we can see that this van der Waals like behaviour occurs whatever the value of PFDM parameter $\alpha$. 

\begin{figure*}
\centering
\includegraphics[scale=0.31]{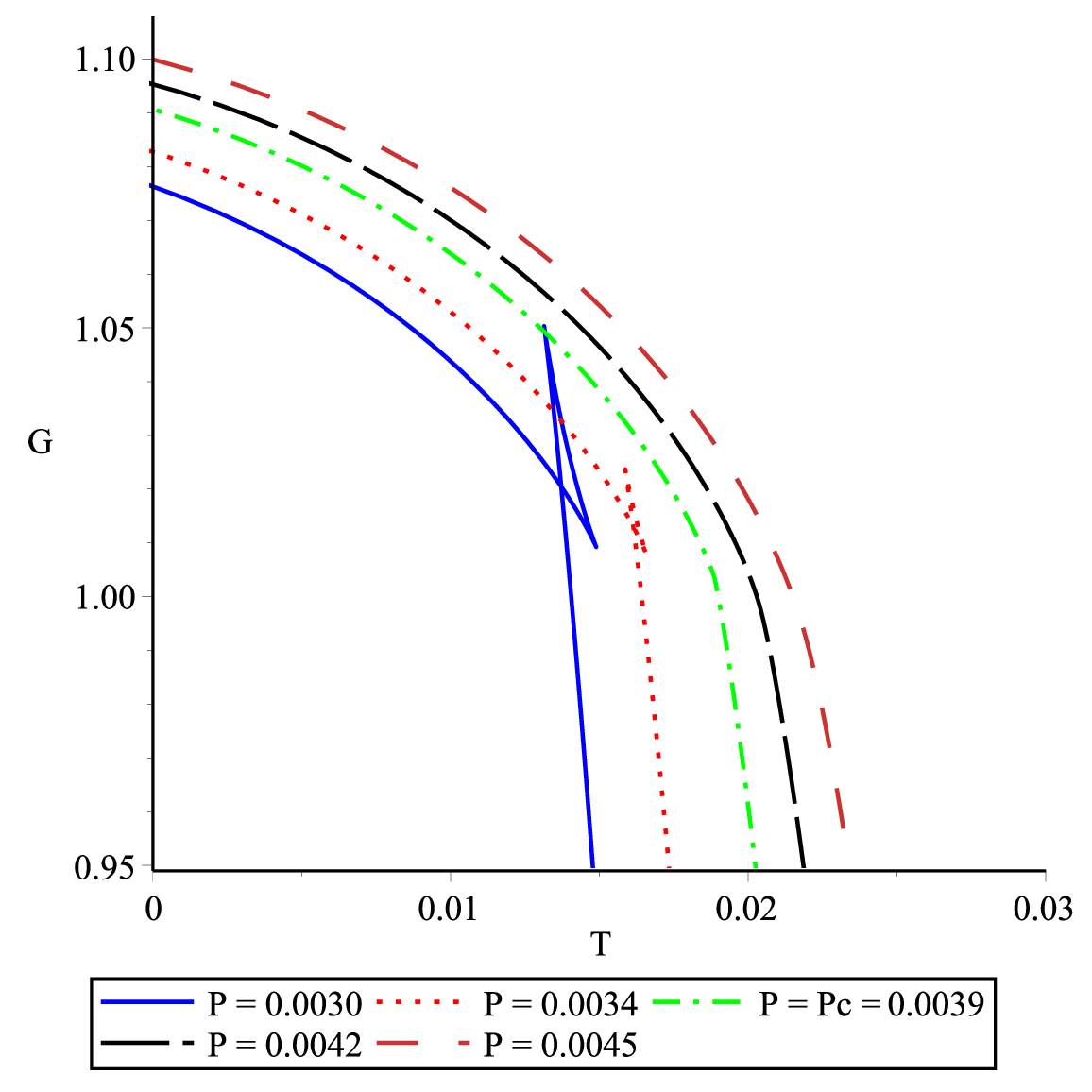}
\caption{\label{fig2}Variation of the Gibbs free energy $G$ for different values of $P$, with $(Q, c_q, \epsilon,\alpha)=(1, 0.2, -2/3,0.4)$.} 
\end{figure*}

On the other hand, one can get the Gibbs free energy $G$, through the following relation
\begin{equation}
G=H-TS,
\end{equation}
where the enthalpy $H$ is considered as the black hole mass. In Fig.~\ref{fig2}, we depicted the behaviour of the Gibbs free energy. Here, we can see that the plot presents the swallow tail characteristic, which appears below the critical temperature and critical pressure(we can see the corresponding critical values for $\alpha=0.4$ on the table~\ref{tab1}), and hence confirms the existence of first order phase transition. Furthermore, in Fig.~\ref{fig2}, the swallow tail behaviour disappears, for higher values of the pressure and temperature. Therefore, we can say that quintessence and PFDM allow the occurrence of a swallow tail on Gibbs free energy behaviour.  

To get more information about phase transition, we have also studied the heat capacity $C$, and see how the presence of PFDM and quintessence impact the behaviour of the black hole. The heat capacity is expressed as 
\begin{equation}\label{heat}
C=T\left(\frac{\partial S}{\partial T}\right)_{Q,c_q,\epsilon,P}.
\end{equation}

Taking into account Eqs. \eqref{T2} and \eqref{entro2},  \eqref{heat}, becomes
\begin{equation}\label{heat2}
C=\frac{2\pi(Q^3+r_h^3)^2}{r_h}\left(\frac{A}{B-D}\right),
\end{equation}

with
\begin{eqnarray}
\begin{array}{r c l}
A&=&-3r_h^{3\epsilon}\alpha Q^3\ln(\frac{r_h}{|\alpha|})+r_h^{3\epsilon}\alpha Q^3-2r_h^{3\epsilon+1}Q^3\\
&+&3c_q\epsilon Q^3+3r_h^3c_q\epsilon+8\pi P r_h^{3\epsilon+6}+3c_qQ^3\\
&+&\alpha r_h^{3\epsilon+3}+r_h^{3\epsilon+4},
\end{array}
\end{eqnarray}

\begin{eqnarray}
\begin{array}{r c l}
B&=&32\pi PQ^3r_h^{3\epsilon+6}+8\pi Pr_h^{3\epsilon+9}\\
&+&6r_h^{3\epsilon}\alpha Q^6\ln(\frac{r_h}{|\alpha|})+15Q^3\alpha r_h^{3\epsilon+3}\ln(\frac{r_h}{|\alpha|})\\
&-&5r_h^{3\epsilon}\alpha Q^6+2Q^6r_h^{3\epsilon+1}-7Q^3\alpha r_h^{3\epsilon+3}, 
\end{array}
\end{eqnarray}

\begin{eqnarray}
\begin{array}{r c l}
D&=&10Q^3 r_h^{3\epsilon+4}-2\alpha r_h^{3\epsilon+6}-r_h^{3\epsilon+7}-9Q^6c_q\epsilon^2\\
&-&18r_h^3Q^3c_q\epsilon^2-9r_h^6c_q\epsilon^2-15Q^6c_q\epsilon\\
&-&21r_h^3Q^3c_q\epsilon-6r_h^6c\epsilon-6Q^6c_q-15r_h^3Q^3c_q.
\end{array}
\end{eqnarray}

\begin{figure*}
\centering
\begin{minipage}[!h]{7cm}
	\centering
	\includegraphics[scale=0.30]{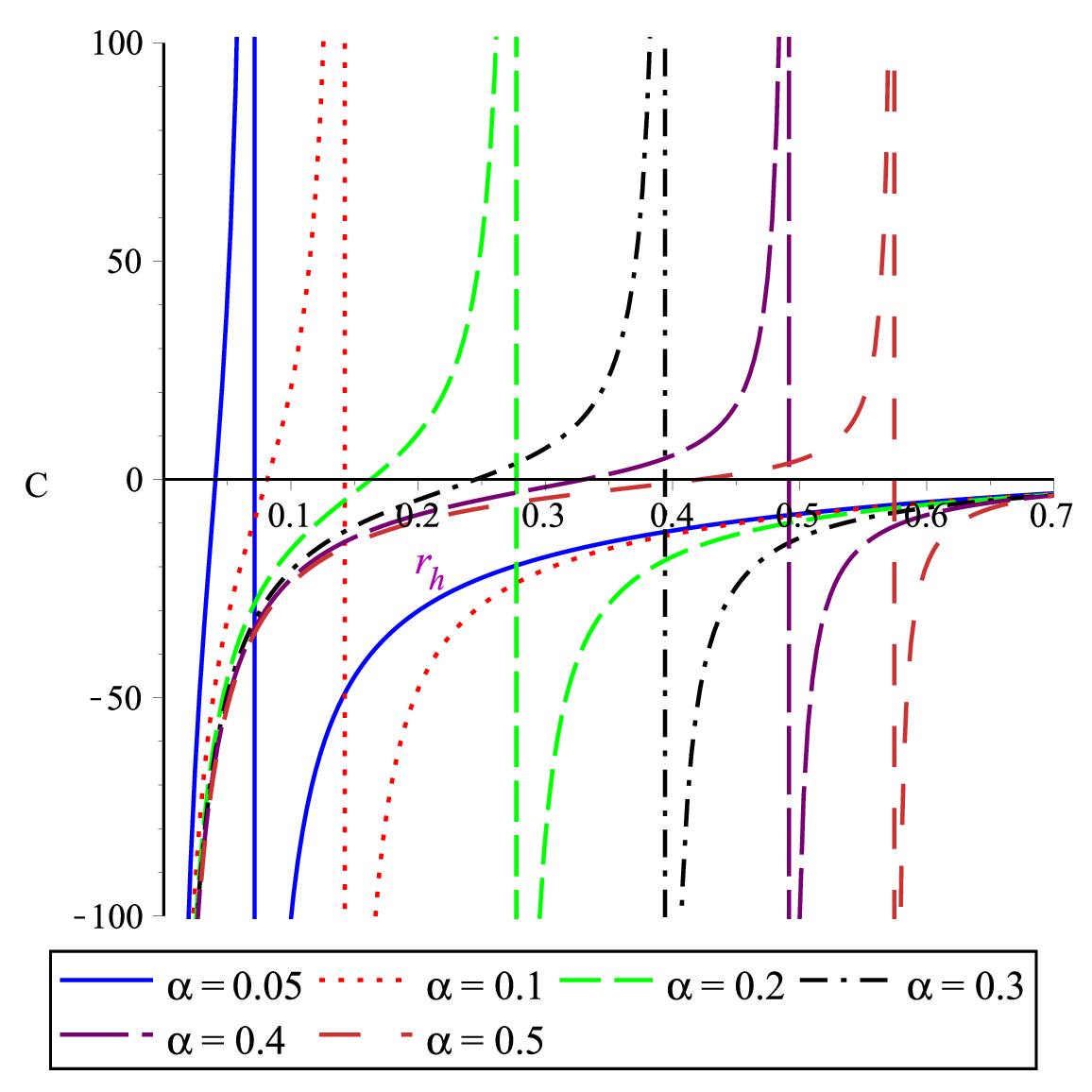}
	
	(a) For smaller values of horizon radius with $(Q, c_q, \epsilon)=(1, 0.02, -2/3).$
\end{minipage}
\begin{minipage}[!h]{7cm}
	\centering
	\includegraphics[scale=0.30]{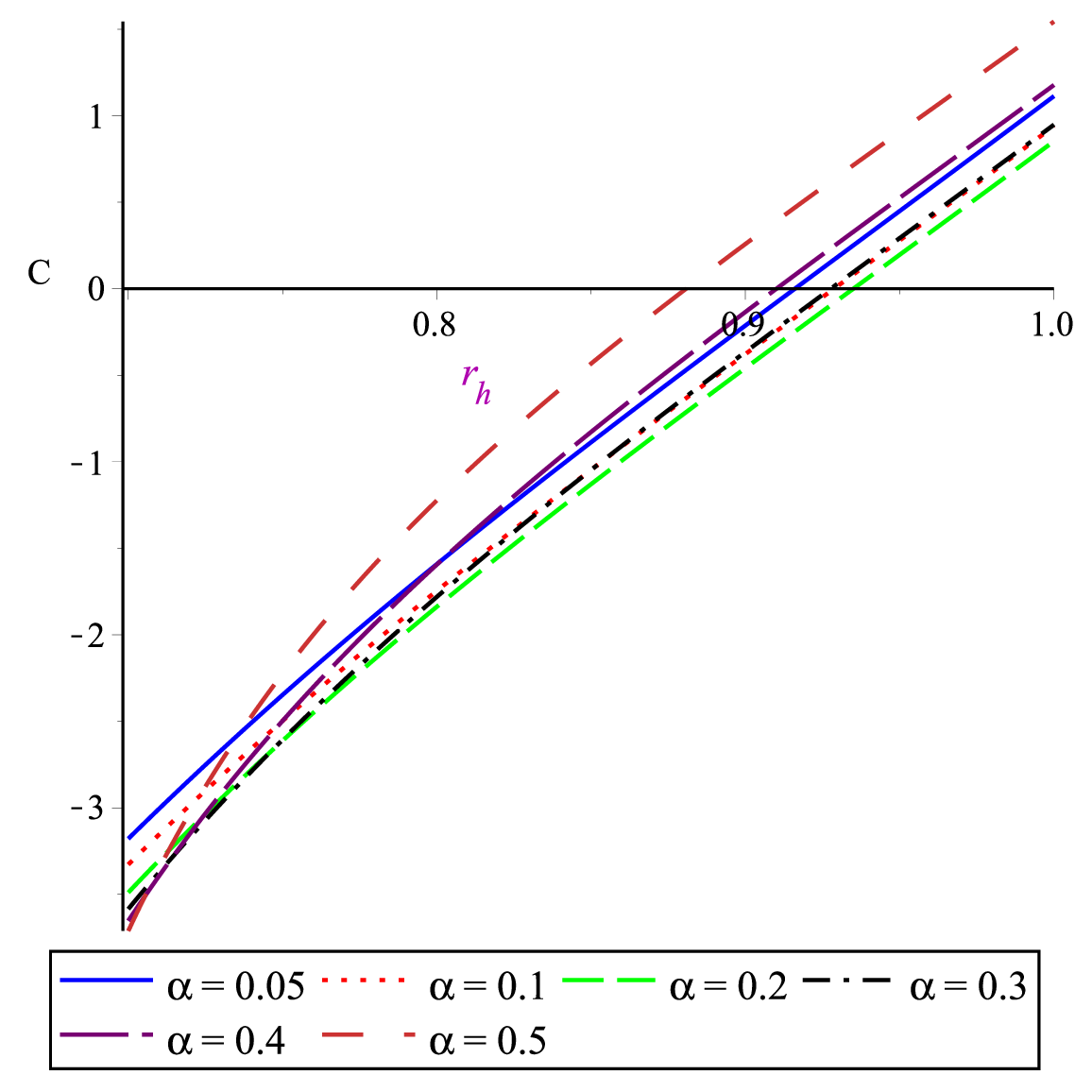}
	
	(b) For higher values of horizon radius with $(Q, c_q, \epsilon)=(1, 0.02, -2/3).$
\end{minipage}
\begin{minipage}[!h]{7cm}
	\centering
	\includegraphics[scale=0.30]{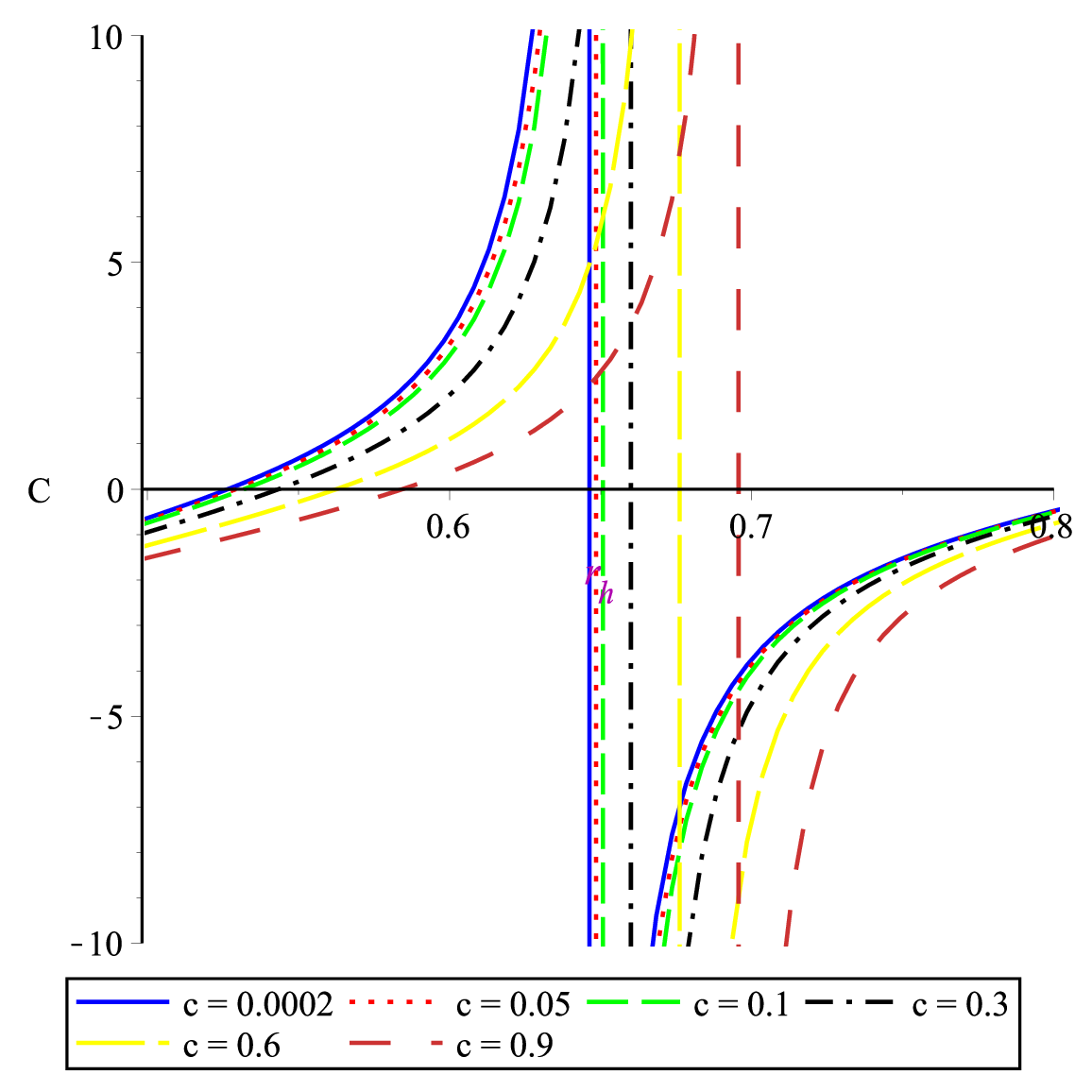}
	
	(c) For smaller values of horizon radius with $(Q, \alpha, \epsilon)=(1, 0.6, -2/3).$
\end{minipage}
\begin{minipage}[!h]{7cm}
	\centering
	\includegraphics[scale=0.30]{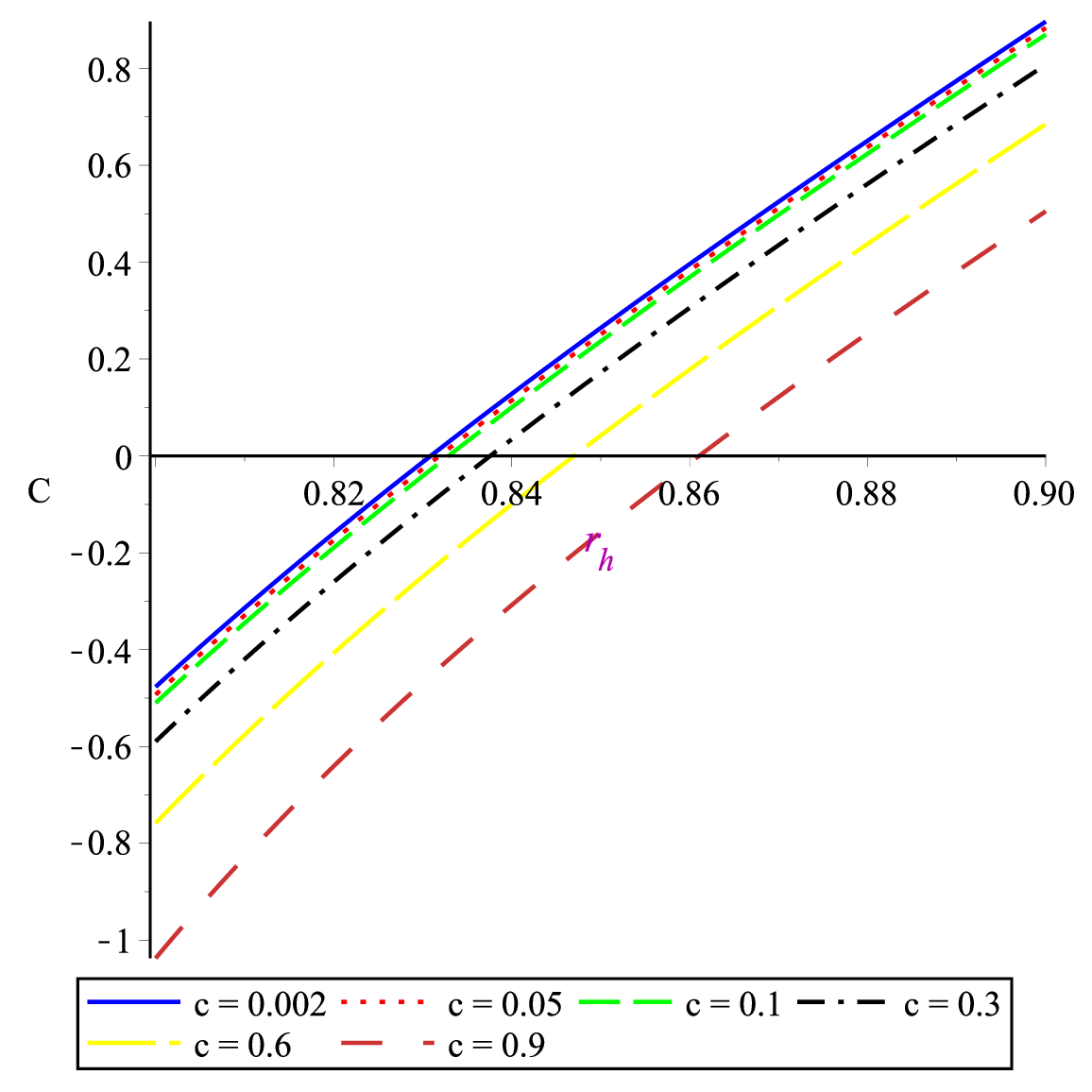}
	
	(d) For higher values of horizon radius with $(Q, \alpha, \epsilon)=(1, 0.6, -2/3).$
\end{minipage}
\caption{\label{fig3}Variation of Heat capacity $C$ in term of horizon radius.} 
\end{figure*}

With this in mind, in Fig.~\ref{fig3} (a) and (b), the heat capacity is reported for different values of PFDM parameter $\alpha$. In these plots, it comes that the heat capacity moves from stable phase to unstable one through a second-order phase transition. Here, the unstable phase is remarked by the negative heat capacity($C<0$), the stable one by the positive heat capacity($C>0$) and the discontinuity on the plot of heat capacity is identified as the second-order phase transition(see Fig.~\ref{fig3} (a) and (c)).  Also, notice that this phase transition is shifted towards higher values of the horizon radius as the PFDM parameter increases(see Fig.~\ref{fig3} (a)). Furthermore, for higher values of  the horizon radius(see Fig.~\ref{fig3} (b)), we can see that the heat capacity of the black hole moves from unstable phase to stable phase without any discontinuity, and hence without second-order phase transition. The same behaviour is noticed before the appearance of the phase transition in Fig.~\ref{fig3} (a).

On the other hand, in other to get more information about the stability of the black hole, in  Fig.~\ref{fig3} (c) and (d), the heat capacity is plotted for different values of the quintessence parameter $c$. Analysing these plots, we observe that a second-order phase transition occurs for lower values of the horizon radius.  Also, notice that this phase transition is shifted towards higher values of the horizon radius as the quintessence parameter increases. Furthermore for higher values of the horizon radius, the black hole moves from unstable phase to stable phase without second-order phase transition(see Fig.~\ref{fig3} (d)). 

However, looking at Fig.~\ref{fig3} (a) and ~\ref{fig3} (c), which correspond to smaller values of the horizon radius, we see that the black hole moves from unstable phase to stable phase, which implies that the black hole to undergo a first-order phase transition, but this could not be the case if for example we take into account quantum corrections such as logarithmic correction as it is done in ~\cite{upadhyay2019modified,ghaffarnejad2022magnetic}. For instance, in ~\cite{ghaffarnejad2022magnetic}, the authors found that the Banados-Teitelboim-Zanelli black hole in massive gravity undergoes a first-order phase transition, but this is not the case when avoiding the correction. Furthermore, another correction such as exponential corrections~\cite{ghaffarnejad2022magnetic} could also lead to a deep change on the behaviour of the heat capacity.

Next, let us consider the relationship between the density of PFDM and quintessence in term of $\alpha$ and $c_q$, respectively. From (\ref{Tuv1}), they are expressed as $\rho_\textmd{PFDM}=-\frac{1}{8\pi}\frac{\alpha}{r^3}$ and $\rho_\textmd{quint}=-\frac{3\epsilon c_q}{2r^{3(\epsilon+1)}}$. Then we study the impact of them on the behaviour of the black hole. These relations clearly show that the second-order phase transition occurs in the presence of both PFDM and quintessence dark energy, and this phase transition is shifted towards higher values of horizon radius as we decrease the PFDM density and increase the quintessence density.


Another way to study the stability of the black hole is through the determinant of Hessian matrix of the black hole mass with respect to its extensive variables $(M(S,Q))$\cite{hendi2015thermodynamic}. In such analysis, the positivity of this determinant also represents the local stability.

Hence, the Hessian matrix is defined as follows:

\begin{eqnarray}
\mathbf{H}^M_{S,Q}=[H_{ij}]=\left[
\begin{array}{r c l}
\frac{\partial^2 M}{\partial S^2}& \frac{\partial^2 M}{\partial S \partial Q}\\
\frac{\partial^2 M}{\partial Q \partial S}& \frac{\partial^2 M}{\partial Q^2}
\end{array}
\right]
\end{eqnarray} 
The compute each of these components we need to have the first derivative, expressed as

\begin{eqnarray}
\begin{array}{r c l}
\frac{\partial M}{\partial S}&=&\frac{\partial M}{\partial r_h}\frac{\partial r_h}{\partial S}\\
&=&3\,{r}^{-3\,\epsilon }{Q}^{3}c\epsilon +3\,{r}^{-3\,\epsilon +3}c
\epsilon +8\,\pi \,P{r}^{6}\\
&+&3\,{r}^{-3\,\epsilon }{Q}^{3}c+3\,\ln \left(  \left| \alpha \right|  \right) {Q}^{3}\alpha\\
&-&3\,\ln  \left( r\right) {Q}^{3}\alpha+{Q}^{3}\alpha-2\,{Q}^{3}r+\alpha\,{r}^{3}\\
&+&{r}^{4}/(4\pi r^2(Q^3+r^3)),
\end{array}
\end{eqnarray}

and

\begin{eqnarray}
\begin{array}{r c l}
\frac{\partial M}{\partial Q}&=&\frac{3Q^2}{2r_h^2}\left(1+\ln\left(\frac{r_h}{|\alpha|}\right)-\frac{c_q}{r_h^{3\epsilon+1}}\right)
\end{array}
\end{eqnarray}.

\begin{figure*}
\centering
\includegraphics[scale=0.31]{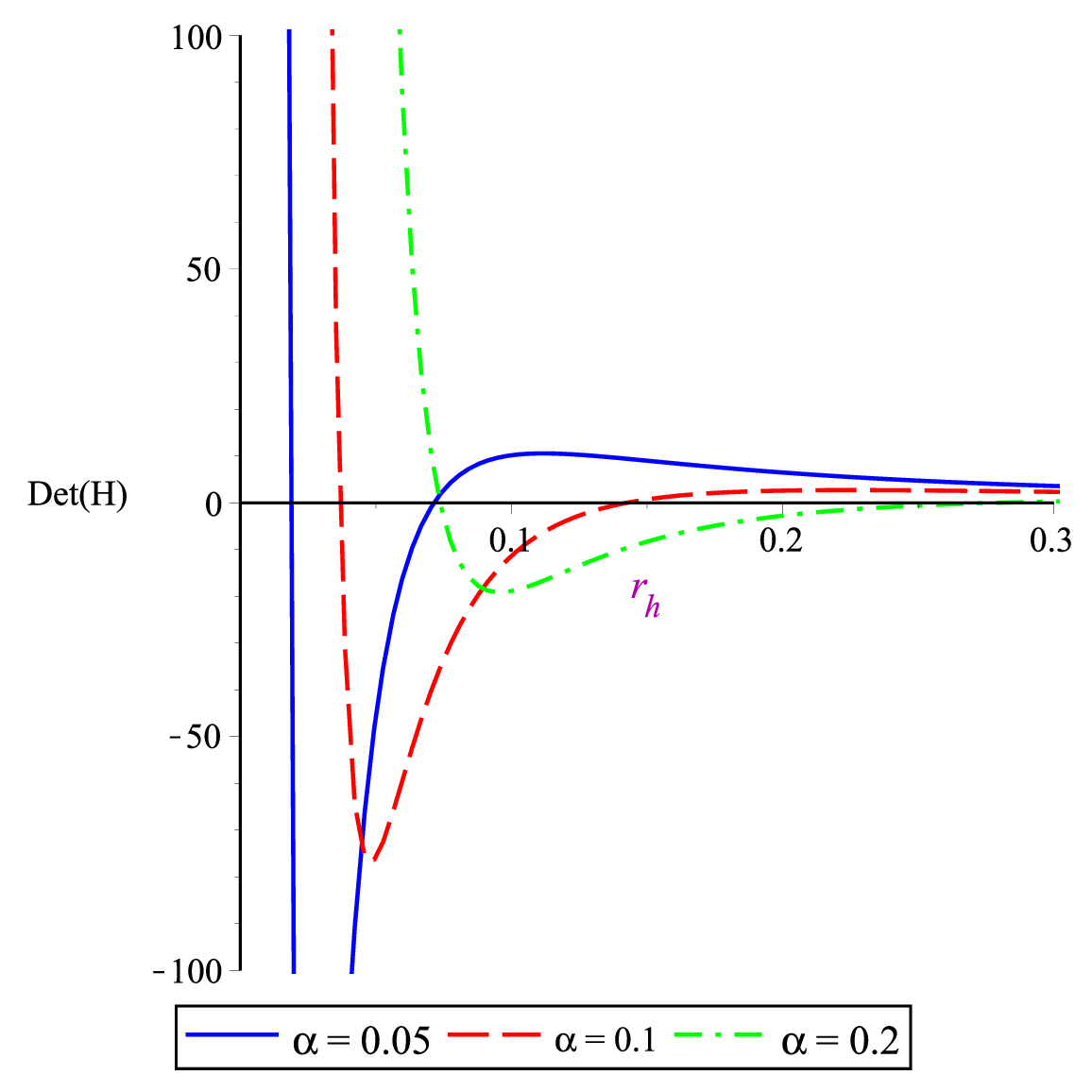}
\caption{\label{hessplot}Variation of the determinant of the Hessian matrix for different values of $\alpha$, with $(Q, c_q, \epsilon,P)=(1, 0.02, -2/3,0.1)$.} 
\end{figure*}

Now, after computing the determinant $Det(\mathbf{H}^M_{S,Q})$ of the Hessian matrix, we plotted it in Fig \eqref{hessplot}, we found that the black hole also has the possibility to be stable, (for $Det(\mathbf{H}^M_{S,Q})>0$ ). Furthermore, we can see that its change of sign appears later as the dark matter parameter $\alpha$ decreases. This results also corresponds to the behaviour of the heat capacity.

\section{Conclusion}
\label{sec:concl}
In summary, we have studied the thermodynamics of non-linear magnetic-charged black hole surrounded by quintessence in the PFDM background. First, we found the corresponding metric, starting by the action, and then deriving the Einstein-Maxwell equations of motion. Using the energy-momentum tensor of quintessence obtained by Kiselev, and the one for PFDM, used by Zhang et \textit{al.}~\cite{zhang2021regular}, the metric we found corresponds to the metric obtained by Ma et \textit{al.}~\cite{ma2021shadow}, for the quintessence parameter $c_q=0$ and cosmological constant $\Lambda=0$.

Secondly, considering the cosmological constant term $\Lambda$ as a dynamical pressure, we found the thermodynamic quantities at the horizon radius, namely the mass, temperature, entropy, pressure and the heat capacity. We found that they are affected by the presence of quintessence and PFDM. In particular, for fixed values of quintessence parameters, the critical values of temperature and pressure increase as we increase the PFDM parameter $\alpha$ (see table~\ref{tab1}). Furthermore, our analysis led us to plot the pressure, with respect to the horizon radius. In this plot, we localized two regions, and the presence of a first-order phase transition, which allows us to move from one region to the second one. Indeed, we found a first region $(T < T_c)$, remarked by one value of radius for a high pressure, and two or three horizon radii for low pressure, and the second region $(T > T_c)$ is remarked by only one horizon radius, for any value of pressure $P$ (see Fig. 1). This behaviour is similar to the van der Waals gas one.

Thirdly, another phase transition we observed is the second-order phase transition, localised in the plot of the heat capacity, with respect to the horizon radius (see Fig.~\ref{fig3}). Looking at this the plot, we saw that as we increase PFDM parameter $\alpha$ as well as the quintessence parameter  $c_q$, the second-order phase transition occurs, and is shifted towards higher values of the horizon radius.

Finally, taking into account the relationship between PFDM and quintessence densities in term of $\alpha$ and $c_q$, respectively, we found that the second-order phase transition is shifted towards higher values of the horizon radius as the PFDM density decreases and the quintessence density increases.


\bibliographystyle{unsrt} 
\bibliography{bibli3} 

\end{document}